# Light mediated emergence of surface patterns in azopolymers at low temperatures

V. Teboul,[1] R. Barillé,[2] P. Tajalli,[3] S. Ahmadi-Kandjani,[3] H. Tajalli,[3] S. Zielinska,[4] and E. Ortyl[4]

[1]*Laboratoire de Photonique d'Angers EA 4464, University of Angers, Physics Department, 2 Bd Lavoisier, 49045 Angers, France.*
[2]*MOLTECH Anjou, UMR CNRS 6200, University of Angers, Physics Department, 2 Bd Lavoisier, 49045 Angers, France.*
[3]*Research Institute for Applied Physics and Astronomy, University of Tabriz, Tabriz, Iran.*
[4]*Wroclaw University of Technology, Faculty of Chemistry, Department of Polymer Engineering and Technology, 50-370 Wroclaw, Poland.*

Polymer thin films doped with azobenzene molecules do have the ability to organize themselves in spontaneous surface relief gratings (SRG) under irradiation with a single polarized beam. To shed some light in this still unexplained phenomenon, we use a new method that permits us to access experimentally the very first steps of the pattern formation process. Decreasing the temperature, we slow down the formation and organization of patterns, due to the large increase of the viscosity and relaxation time of the azopolymer. As a result decreasing the temperature allows us to access and study much shorter time scales, in the physical mechanisms underlying the pattern formation, than previously reported. We find that the patterns organize themselves in sub-structures which size increase with the temperature, following the diffusion coefficient evolution of the material. That result suggests that the pattern formation and organization is mainly governed by diffusive processes, in agreement with some theories of the SRG formation. Decreasing further the temperature we observe the emergence of small voids located at the junction of the sub-structures.

**INTRODUCTION**

When doped with azobenzene molecules, amorphous materials under irradiation are subject to intriguing massive mass transport [1]. These still unexplained molecular motions lead eventually to the spontaneous organization of the material and to the formation of surface relief gratings (SRG). While the physical mechanisms leading to these mass transport and to the material organization are the object of conjectures [2-13], there is no doubt that it originates from the photo-isomerization property of the azobenzene molecule. Various mechanisms have been proposed [1] to explain that transport. The proposed mechanisms include the mean field induced by the dipolar attraction between the azo-dyes [3,4], the incident light electric field gradient 5, the mechanical stress induced by the orientation of the dyes [6], the pressure gradients induced inside the material by the isomerizations [7,8], an isomerization-induced cage breaking process around the azo-dyes [13], then followed (or not) by the modification of spontaneous cooperative mechanisms in soft matter [2,12], the periodic modification of the free-volume [9,10] around the dye induced by the isomerization, and the reputation of the dye along the polarization direction [11]. Interestingly, a few important results recently shed some light in that still unexplained huge isomerization induced transport. First, a fluidization of the medium in the vicinity of the chromophore has been reported experimentally by different groups [14-16]. Another important result is that the pressure needed to stop the isomerization of the chromophores is very high (P > 1GPa) [17]. Finally, while the SRG organization of the material is usually obtained from an interference pattern, with the matter moving in the direction of the dark zones, it has been shown experimentally18{20 that even with a single polarized beam, the SRG organization appears. Because there are no interference fringes to guide the organization of the medium when a single beam is used, the physical mechanism has to include a self-organization of the
medium [18 -20] in that experiment. This phenomenon has been interpreted21 as arising from the interplay of the interferences between light radiated from defects in the sample, and the photo-isomerization induced local density modifications resulting from these interferences. In this work, with that single beam experiment, we use a new method that permits us to access the very first steps of the patterns formation and organization process. Below their melting temperature, amorphous materials display a universally large increase [22 - 24] of their viscosity ν when their temperature drops: $\nu = \nu_0 e^{E_a/k_B T}$ where $E_a(T)$ is an activation energy and is a constant. That increase of the viscosity hinders molecular motions at low temperatures. As a result using a constant experiment duration, a temperature drop corresponds to a shorter effective time for the physical mechanisms that create the patterns. We thus use the large increase of the viscosity and relaxation time of amorphous materials when the temperature drops, to slow down the organization of patterns and to access much shorter time scales, in the physical mechanisms underlying the patterns formation and organization, than previously reported. We find that the patterns organize themselves in sub-structures which size increase with the temperature, following the diffusion co-efficient evolution of the material. That result suggests that the patterns organization is mainly governed by diffusive processes, in agreement with some theories of the SRG formation. Decreasing further the temperature we observe the emergence of new and strange nano honey-comb voids located at the junction of the sub-structures. We then use a different method to increase the viscosity of our samples. We decrease the thickness of the sample leading to a confinement effect [25 - 33] that results in a modification of the viscosity. We observe very similar patterns with confinement at room temperature than with a temperature decrease without confinement.

**EXPERIMENTAL SECTION**

In order to study the temperature dependence of the surface grating formation, the thin films were positioned on a computer-controlled, closed-cycle helium refrigeration (CTI Cryogenic, model 22) temperature chamber with a temperature uncertainty of 0.5 K. Spontaneous patterns were inscribed using an Ar+ ion laser operating at a blue wavelength of 488 nm with linear s-polarization. This wavelength is close to the absorption maximum of the azobenzene dye. The laser beam was expanded and spatially selected in order to use only its homogeneous central part. To avoid the air turbulence, the experiment was performed inside a low pressure chamber. Lately the morphology of induced patterns was analyzed using an atomic force microscope (AFM) with contact mode. Polymer thin _lms are made from a highly photoactive azobenzene derivative containing heterocyclic sulfonamide moieties (IZO1). The details of synthesis of 2-[4-[(E)-(4-[(2,6-dimethyl pyrimidin 4 - yl)amino] sulfonylphenyl) diazenyl]phenyl- (methyl)amino]ethyl2-methylacrylate are reported elsewhere35. The monomers of the methylacrylate type contain aliphatic spacers of di_erent length between chromophoric and methylacrylic groups. The molecular mass of the polymer determined by GPC was found in between 14000 g/mol and 19000 g/mol, and the glass transition temperature of the polymer is Tg = 344K. Note however that the free surface of the polymer film reduces the viscosity of the sample leading to a smaller Tg value around the surface. Our thin films were prepared by dissolving 75 mg of azopolymer in 1 ml of THF (Tetrahydrofuran) and spin-coated on a clean glass substrate. The _lm thickness was then determined with a Dektak profilometer and was found around $e$ = 550 nm. Films of IZO1 azopolymer with 550 nm thickness were irradiated during 4 hours with a linearly polarized argon ion laser beam at 488 nm with a vertical polarization. The laser beam impinges on the sample with normal incidence. We use a relatively small incident beam intensity I = 100 mW/cm$^2$. The temperature of the sample was varied from 298 K to 73 K. The samples stayed in the cryogenic chamber during 2 hours for thermalization before illumination by the laser. The thermalization procedure is expected to stabilize the temperature and to eliminate any remaining stress inside the sample. Moreover the isomerizations of the dyes that take place during the illumination and create the patterns, have been found to induce a local decrease of the viscosity that would overcome any remaining localized stress. We didn't find any noticeable surface structure before the samples illumination. We chose a 4 hours irradiation time as this is the time needed to obtain a grating when T = 73K (the smallest temperature studied in our work). Every photoinduced grating used in this work corresponds to a different sample. We then analyze the mean surface of organized zones in the samples AFM pictures. For that purpose we use an image-processing software and we process

the surface-pattern images through a Gaussian low pass filter to reduce the amplitude of non periodic
grating details and irregularities by averaging. This step separates the areas inside the patterns with or without organized structures and helps enhancing the organized areas. Afterwards, for each temperature, we prepare sets of contour domains by thresholding at different intensity levels the pattern amplitude of the image modified by the filter. This results in outline drawings consisting of
non-intersecting closed level zones, and each of these contours is then filled to create black and white image masks. Then we calculate the average surface of all these new domains, and then normalize that surface with the whole sample organized surface.

**RESULTS AND DISCUSSION**

After irradiation of the azopolymer thin films, we analyze the different samples with an AFM microscope using the contact mode. The resulting pictures display different surface patterns depending on the temperature of the sample. We study two different temperature ranges.
The first temperature range starts at room temperature for which previous results exist[36] down to 273 K, and the second temperature range starts around 273 K down to 73 K. We display the results of that second set of experiments in Figure 1.

Figure 1

The AFM pictures in Figure 1 show self-structured SRG patterns. For the largest temperature that we display (T = 273K, $T/T_g$ = 0.79) the patterns organization extends to the whole picture surface. This is roughly the same picture that we also observe at larger temperatures.
That result shows that the timescale of the experiment is sufficient to reach the equilibrium and the gratings are complete. When we decrease the temperature down to T = 173K ($T/T_g$ = 0.50, second picture) a few patterns of smaller sizes and aligned in different directions appear. Decreasing further the temperature we find that the small size unaligned patterns progressively increase in number, eventually extending to the whole sample surface at the temperature T = 73K. The Fourier transform pictures similarly display the progressive disorganization of the medium when the temperature decreases, with an increasing number of spots. As argued in the introduction, a temperature drop corresponds to an increase of the relaxation times of the material, and as the duration of the experiment is set constant, a temperature drop is equivalent to a decrease of the experimental observation time. Using that viewpoint, we interpret the pictures of Figure 1 as a movie of the patterns formation, reading the picture from the right to the left hand side. Our results thus show that the material surface first organizes itself in small unaligned patterns. Then some larger patterns appear aligned in a particular direction. Eventually these larger aligned patterns merge creating the surface relief grating. The similarity between the small unaligned patterns and the large aligned patterns, together with the progressive evolution that we observe, confirms us that no other physical effects (for example partial crystallizations of the polymer) than the SRG formation are present in our experiments. In the first picture from the right, i.e. for the shorter times or smaller temperatures studied, the different pat- terns display two roughly perpendicular alignments, none of them being in the direction finally chosen by the system for the SRG as shown by comparison with the first picture from the left. The pattern directions are thus not totally disorganized in the pictures as every pattern direction is not present. The beam polarization that has to be at the origin of that effect is parallel to the white lines on the Figure, and perpendicular to the final patterns. We observe on that picture a mosaic of zones of aligned patterns, each zone being aligned in a different direction from the zones surrounding it. Interestingly the surfaces of the different aligned zones are roughly constant in the picture.

Figure 2

A zoom of the AFM pictures for the lowest investigated temperatures, shows in Figure 2 the appearance of small honeycomb voids located at the frontiers of the organized domains. We explain these geometric structures (the voids) as being induced around chromophores by repeated isomerizations. In that picture the repeated isomerizations of the chromophore induce a flow of the surrounding matter that eventually results in a matter depletion in the chromophore vicinity. These hexagonal voids appear for the lowest temperatures investigated, suggesting that this process is one of the first steps in the material organization. Then the matter migrates due to the isomerization-induced diffusion [12] creating the patterns. The specific location of the hexagonal structures at the frontiers of the pattern areas supports that picture of these structures as being one of the first steps in the material organization.

Figure 3

From the whole set of digitalized AFM pictures we then calculate numerically the mean surface of the aligned patterns zones for the different temperatures studied. We display these results in Figure 3. From room temperature to T ≈ 200K the mean surface of the aligned pattern zones follow an exponential evolution with temperature (green dashed line). Then below 200K the exponential evolution slows down tending to a constant low temperature limit < S > ≈ 1.5 µm². That evolution under light exposure is reminiscent of the diffusion co-efficient evolution with temperature when the material is illuminated reported from molecular dynamics simulations in several papers [12, 37, 38]. When the material is not illuminated the diffusion coefficient D follows a similar Arrhenius law than the inverse of the viscosity η as these two quantities are related with the Stokes-Einstein law: $D = k_BT/6\pi\eta a$. In this formula $a$ is the characteristic size of the particles. Note that the $k_BT$ co-efficient linear evolution with T is negligible in front of the exponential behavior of the viscosity. Thus without illumination the diffusion coefficient follows the typical law: $D = D^{thermal} = D_0.exp(-Ea/k_BT)$. Then upon illumination the azo-dyes isomerize leading to an isomerization-induced diffusion that has to be added to the previous thermally activated diffusion, leading to: $D = D_0.exp(-Ea/k_BT) + D^{induced}$. At high temperatures the first term $D^{thermal}$ predominates leading to an exponential evolution, while at low temperatures the diffusion finally tends to $D^{induced}$ as $D^{thermal}$ tends to zero. That similarity between the evolution of the diffusion co-efficient and of the surface of the aligned zones suggests that the increase in surface of the aligned zones is mainly governed by diffusive processes.

Figure 4

Figure 4 shows that the mean pitch (i.e. the patterns groove spacing) follows a linear evolution with the temperature i.e. a logarithm evolution with time, while the zone surfaces in Figure 3 follows an exponential evolution with temperature. The pitches increase by less than 10 percents in the temperature range studied while the surface of the zones is multiplied by a rough factor 4. The organized zones thus increase in size much more rapidly than the pitches. Similarly the patterns height evolution with temperature is also quite small in our experiments. The height of the patterns reaches approximately its final size at the lowest temperature studied (73 K) before the patterns alignments. Thus, the alignment of the patterns follows their formation in our experiments, a result that may be understood from two different pictures, namely that the pattern formation is much more rapid than the alignment or that the alignment needs strong patterns to be efficient.

Figure 5

Using the viscosity (or equivalently the relaxation time) exponential increase with temperature of the sample material, we have interpreted the temperature evolution of the grating as equivalent to its time evolution upon light exposure. That picture is suggested by the time-temperature superposition principle, a principle that is well established for amorphous materials. It appears interesting anyway to make direct measurements of the time evolution of the gratings at constant temperature. We show the results of that time evolution study at room temperature in Figure 5. The disorganization effect is less visible than at low temperature due to the increase of the thermal diffusion. However we observe in Figure 5 an organization of the medium with time.

Figure 6

We will now investigate a different route to increase the viscosity of the sample without changing its temperature. The proximity of a wall is known to increase the viscosity [25 - 33] of the surrounding amorphous material. That effect known as confinement depends on the wall surface roughness [34] and has been the subject of a number of studies due to possible applications in nanotechnology and theoretical implications on the glass-transition problem. Similarly the presence of a free surface decreases the viscosity of the material. In our study, the light impinges on the free surface and then, due to absorption, the number of photo-isomerizations per volume element decreases when we go deeper inside the sample. As a result decreasing the sample thickness we are able to increase the effect of the wall (the substrate) resulting in an increase of the sample viscosity.

We show in Figure 6 the patterns that we obtain with that method at room temperature (T = 300K) but for a sample thickness $e$ = 300nm (0.3 µm) instead of 550nm as in the previous measurements. The Figure shows a disorganisation very similar to what we found with a temperature decrease. That result suggests that the disorganisation is due for both experiments to the increase of the viscosity of the sample. Figure 6 thus confirms our previous interpretation.

**Conclusion**

In this work we have used a new method that permits us to access experimentally the very first steps of the patterns formation and organization process. Decreasing the temperature, we slow down the formation and organization of patterns, using the large increase of the viscosity and relaxation time of the azopolymer. We found that the patterns organize themselves in sub-structures which size increase with the temperature, following the diffusion coefficient evolution of the material. That result suggests that the patterns organization is mainly governed by diffusive processes, in agreement with some theories of the SRG formation. Decreasing further the temperature we observe the emergence of nano honeycomb voids located at the junction of the sub-structures. We interpret these voids as the effect of the successive isomerizations of the chromophores on the surrounding material, in agreement with the fluidization process recently reported around the chromophores14{16 and with the increase of the diffusion predicted by simulations [12].

**Figure caption**

Figure 1: (color online) AFM pictures of the irradiated samples for various temperatures. From the left to the right hand side we have T=Tg = 0:79; 0:50; 0:36; and 0:21 ( T = 273K; 173K; 123K; and 73K). Tg = 71 degrees Celsius (i.e. 344K). The Fourier transform of these pictures is plotted for each pattern showing the distribution of spatial frequencies and the decreasing degree of organization from the left to the right hand side. The laser beam polarization is vertical and the incidence angle is normal to the sample. The polarization direction is parallel to the small white lines.

Figure 2: (color online) Zoom of AFM pictures at low temperatures T = -150C (T = 123K, left) and T = 200C (T = 73K, right). The Figure shows that small voids appear at these low temperatures.

Figure 3: (color online) Average surface of the patterns zones versus temperature. The decrease of the surface follows an exponential law (green line) at high temperatures and then slows down at low temperatures (blue line). This evolution reminds the evolution found for the diffusion coefficient (and the inverse of the $\alpha$ relaxation time) in simulations [12, 37, 38].

Figure 4: (color online) Average pitch of the SRG patterns as a function of temperature. The evolution of the pitch is relatively small. Note that the pitch tends to the theoretical value [21] $\Lambda$ = 976 nm.

Figure 5: (color online) Evolution of the pitch and height of the grating pattern with time, together with AFM pictures of the irradiated sample at room temperature for three different irradiation times. While the effect is less visible at room temperature due to larger thermal diffusion, we also observe an organization of the medium with time.

Figure 6: (color online) AFM pictures of the irradiated sample at room temperature but with a smaller sample thickness $e$ = 0.3 $\mu$m inducing due to confinement an increase of viscosity similar to the effect of a temperature drop [25, 28-33]. The sample was irradiated during 2 hours with an incident beam intensity I = 500mW/cm$^2$. The Fourier transform of the picture is plotted on the right showing the distribution of spatial frequencies and the degree of organization. The small arrow shows the polarization direction.

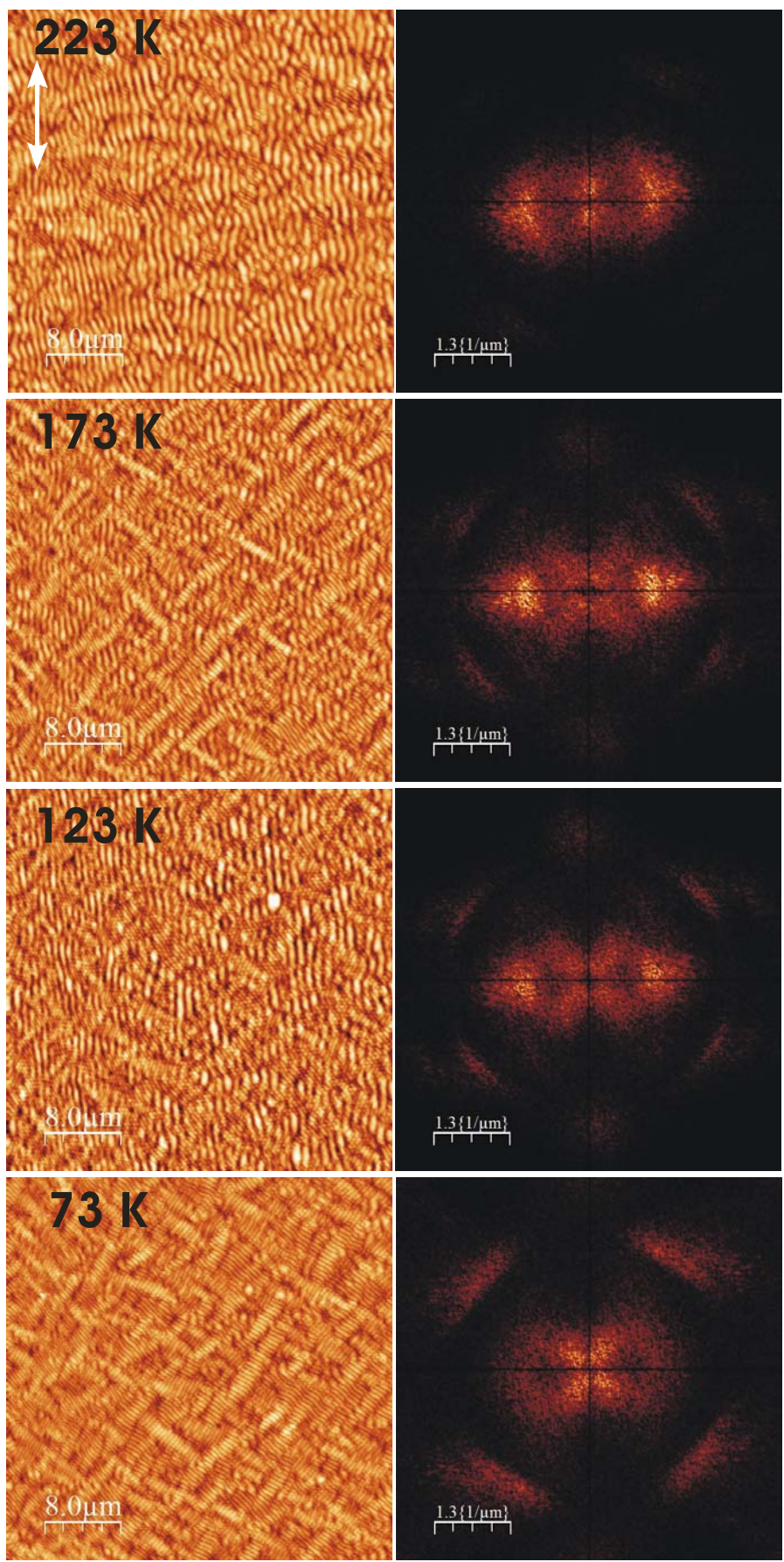

**Figure 1**

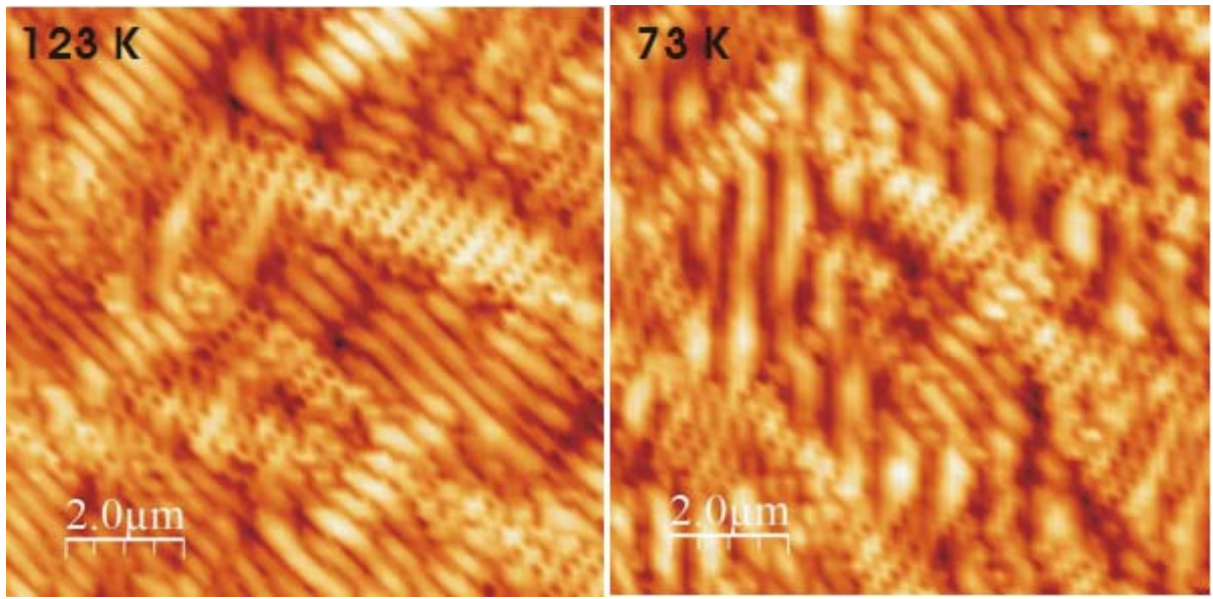

**Figure 2**

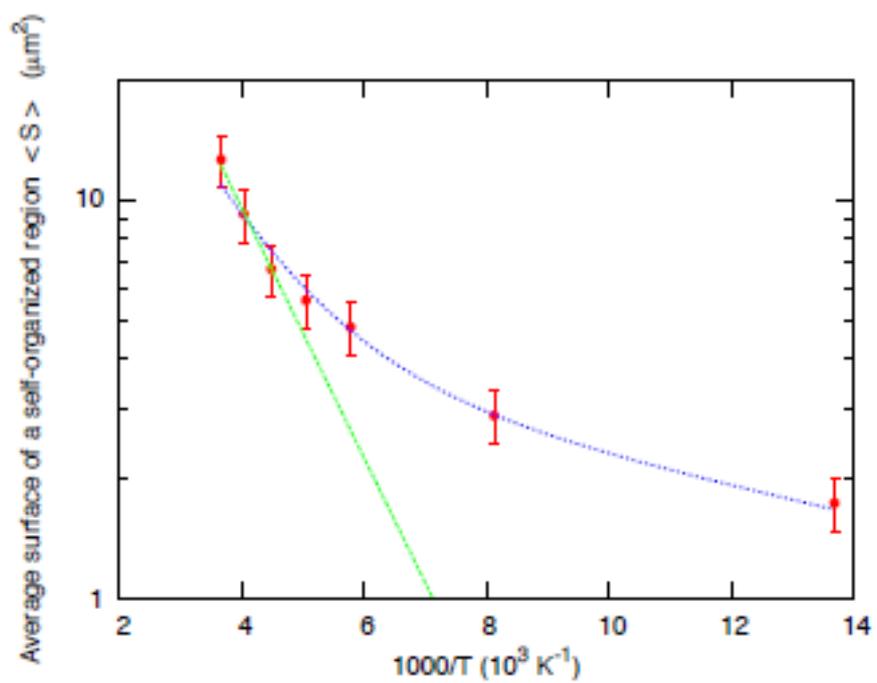

**Figure 3**

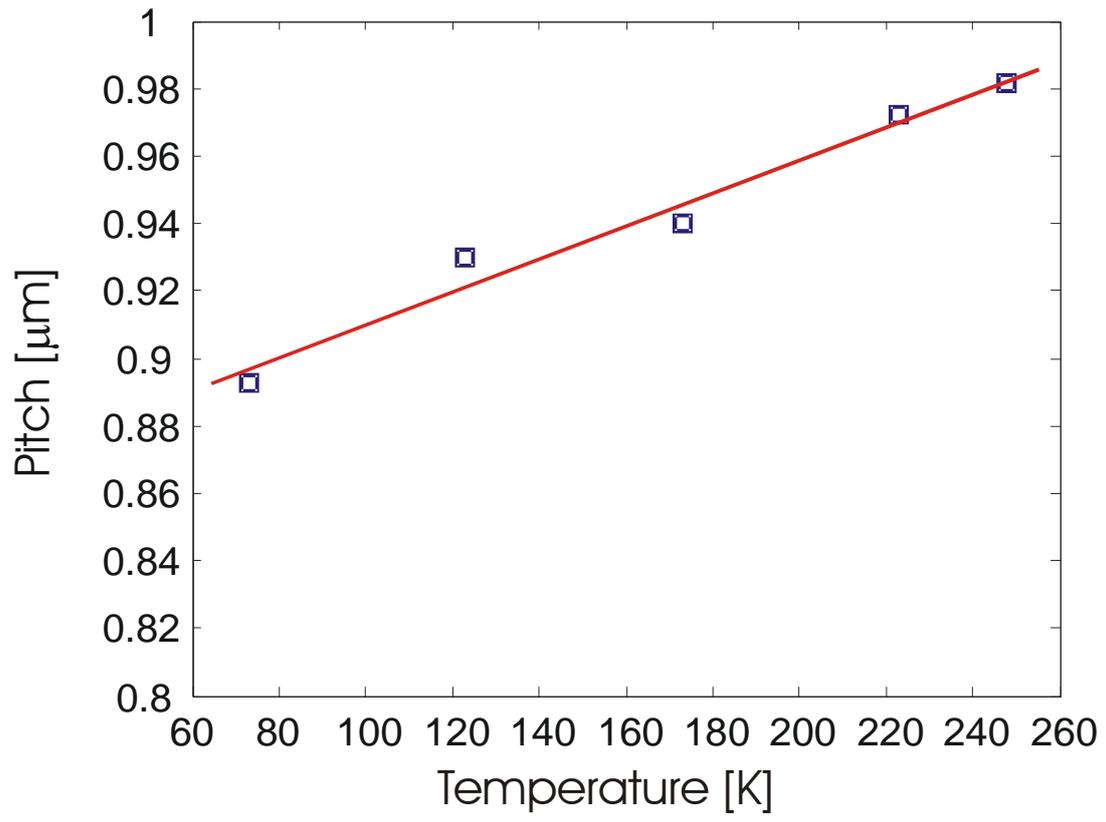

**Figure 4**

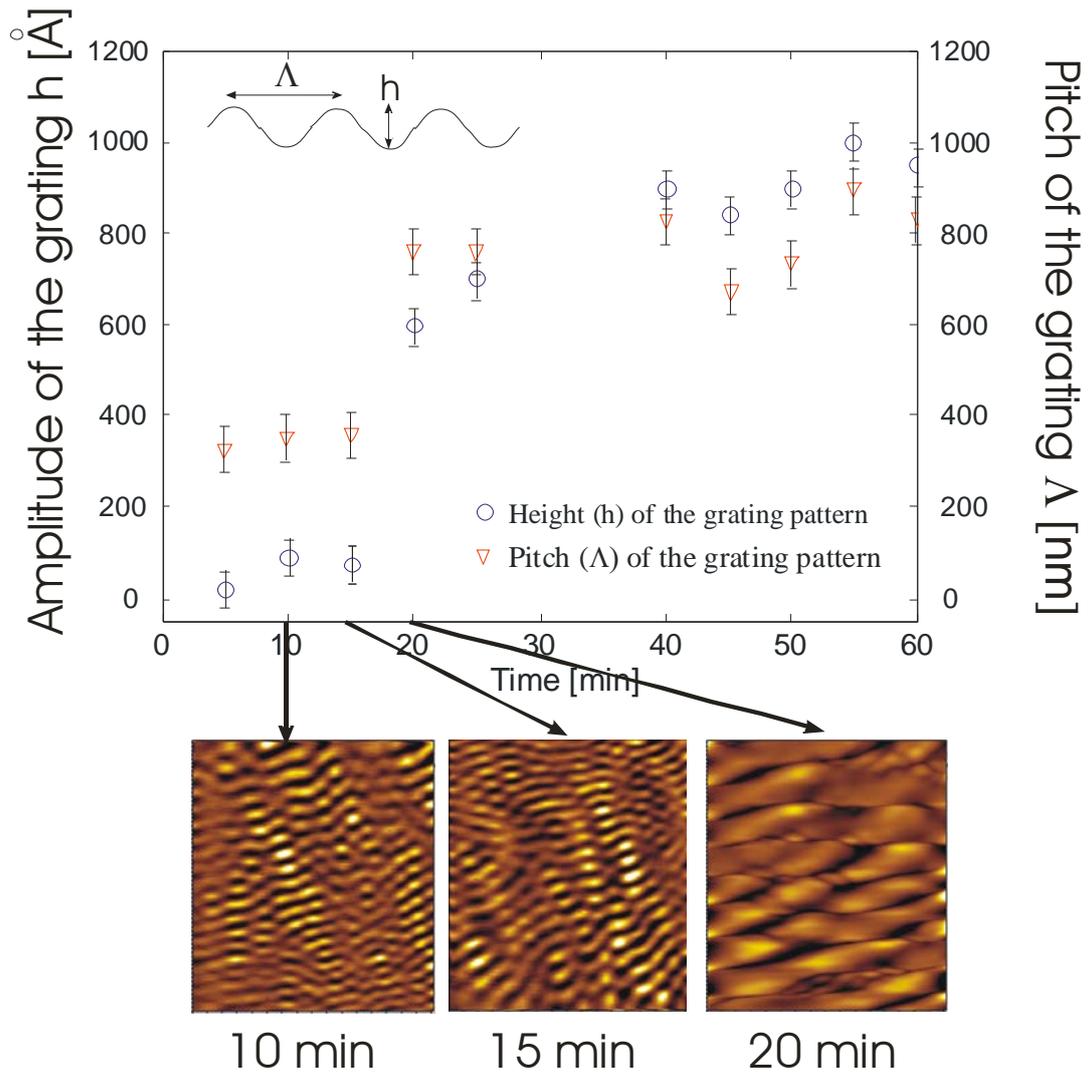

Figure 5

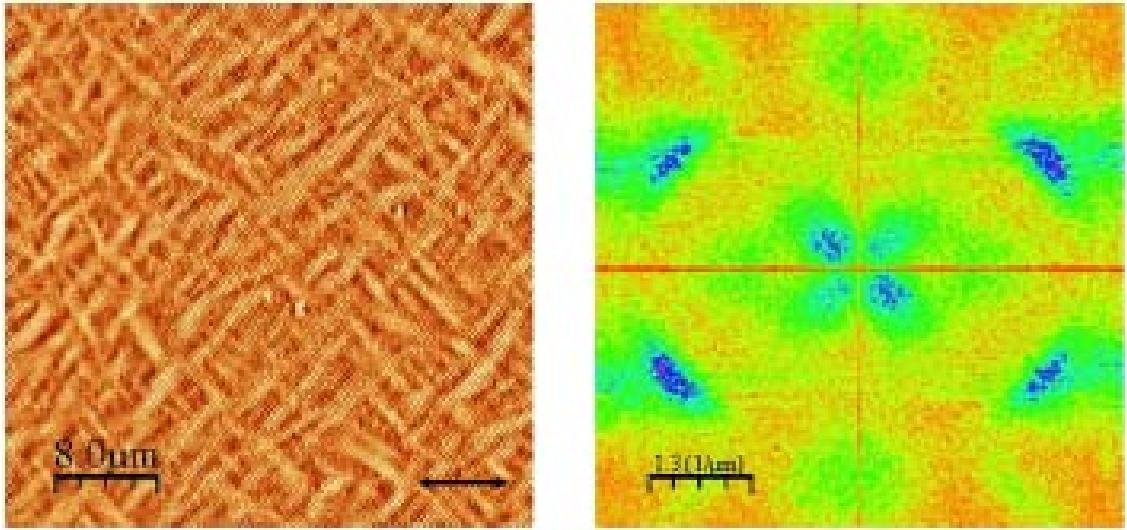

**Figure 6**